\documentclass[12 pt]{iopart}
\usepackage{epsfig}

\begin{document}
\newcommand{\be}{\begin{equation}}
\newcommand{\ee}{\end{equation}}
\newcommand{\ba}{\begin{eqnarray}}
\newcommand{\ea}{\end{eqnarray}}
\newcommand{\Gam}{\Gamma[\varphi]}
\newcommand{\Gamm}{\Gamma[\varphi,\Theta]}
\thispagestyle{empty}
\title[Application of the Fr\"{o}benius method ]{Application of the Fr\"{o}benius method \\
to the Schr\"{o}dinger equation for a spherically symmetric
potential: anharmonic oscillator}
\author{Przemys\l aw Ko\'scik and Anna Okopi\'nska\\
Institute of Physics, Pedagogical University,\\
\'Swi\c{e}tokrzyska 15, 25-406 Kielce, Poland}

\ead{koscik@pu.kielce.pl}

\begin{abstract}

\noindent The power series method has been adapted to compute the
spectrum of the Schr\"{o}dinger equation for central potential of
the form $V(r)={d_{-2}\over r^2}+{d_{-1}\over r}+\sum_{i=0}^{\infty}
d_{i}r^i$. The bound-state energies are given as zeros of a
calculable function, if the potential is confined in a spherical
box. For an unconfined potential the interval bounding the energy
eigenvalues can be determined in a similar way with an arbitrarily
chosen precision. The very accurate results for various spherically
symmetric anharmonic potentials are presented.
\end{abstract}
\maketitle
\section{Introduction}
The exact solution of the Schr\"{o}dinger equation can be obtained
only for a few particular forms of potentials, in other cases one
has to resort to approximations or numerical techniques. Many
approximation methods have been developed for solving problems in
one-dimensional space. Approximate solutions to the Schr\"{o}dinger
equation have been also studied for spherically symmetric potentials
in $D-$dimensional space, both by methods elaborated for
one-dimensional space, e.g. the Hill determinant method
\cite{chaud}, the variational approach \cite{appr}, and by methods
dedicated to D-dimensional problems, e.g. the shifted 1/D expansion
\cite{hel1,hel}. Here we show that highly accurate solutions to the
Schr\"{o}dinger equation can be determined for various types of
spherically symmetric potentials with the use of the Fr\"{o}benius
method (FM). The method consists in expanding the solution of a
differential equation into power series \cite {Fuchs}, and was
originally applied by Barakat and Rosner~\cite{bar} to compute the
spectrum of one-dimensional quartic oscillator confined by
impenetrable walls at $x= \pm R$. The energy eigenvalues of the
system have been obtained numerically as zeros of a function,
calculated from its power series representation. Moreover, it has
been shown that the bound-state energies of the confined system
approach rapidly those of the unconfined oscillator for increasing
$R$. Low-lying eigenvalues for other one-dimensional
potentials~\cite{alh1} have been also successfully calculated in a
similar way. Recently, a modified treatment of unconfined systems
allowed for a very accurate determination of the ground-state energy
for the quartic oscillator~\cite{trott}. In all the cases studied
the potential was a finite function, and a solution was expanded
around an ordinary point of the differential equation. Here we study
the application of the FM for solving the radial Schr\"{o}dinger
equation, which requires that an expansion around a regular singular
point should be used.

The outline of the present work is as follows. In Section~\ref{exp}
the solution of the radial Schr\"{o}dinger equation in form of a
generalized power series is discussed. The case of a spherically
symmetric potential bounded by an impenetrable wall at $r=R$ is
studied in Section~\ref{conf}. In this case, the energy eigenvalues
can be easily determined by finding the roots of the polynomial,
which is illustrated on the example of the confined harmonic and
anharmonic oscillators and Hulth\'{e}n potential. The case of
unconfined system is studied in Section~\ref{meth}, where a  scheme
for determining an arbitrarily large set of bound-state energies is
developed. After demonstrating the performance of the method on the
exactly solvable example of the Kratzer potential, the results for
the unconfined oscillator are presented for various choices of
anharmonic parameters.

\section{Expansion around a regular-singular point}\label{exp}

The Schr\"{o}dinger equation for a spherically symmetric potential
in $3-$dimensional space can be reduced to an ordinary differential
equation in the radial variable \be [-{1\over 2r }{d^2\over
dr^2}r+{l(l+1)\over 2r^2}+V(r)]R(r)=\lambda R(r),\label{schr}\ee
where  $l$ is the angular momentum quantum number, and the units
$\hbar=1,~m=1$ are used. Upon introducing the function $u(r)=rR(r)$,
the differential equation (\ref{schr}) takes the form of the
one-dimensional Schr\"{o}dinger eigenvalue problem \be
\left[-{1\over 2 }{d^2\over dr^2}+V_{eff}(r,l)\right]u(r)=\lambda
u(r),\label{ham}\ee where the effective potential reads \be
V_{eff}(r,l)={l(l+1)\over 2r^2}+V(r).\ee The point $r=0$ is a
regular singular point of the radial equation, if the potential
$V(r)$ diverges but $r^2 V(r)$ remains finite as $r\rightarrow 0$,
which is the case for the interaction potential of the form
 \be V(r)={d_{-2}\over r^2}+{d_{-1}\over
r}+V_{reg}(r),\label{pot}\ee where the regular part is represented
by a convergent series \be V_{reg}(r)=\sum_{i=0}^{\infty}
d_{i}r^i.\label{pot1}\ee In this case the FM can be applied with the
radial wave function represented as a generalized power series
 \be u(r)=r^\delta \sum_{i=0}^{\infty}
a_{i}r^{i}\label{fun},\ee where $a_{0}\neq 0.$  In what follows, we
will take $a_{0}=1$, since normalization of wave function is
irrelevant in our calculation. Substituting (\ref{fun}) into
(\ref{ham}) we obtain the equation
 \be -{1\over 2} \sum_{i=0}^{\infty}
[(i+\delta)(i+\delta-1)-l(l+1)]a_{i}r^{i}
 +(\sum_{i=0}^\infty d_{i-2}r^i )( \sum_{i=0}^{\infty} a_{i}r^{i})=
 \lambda \sum_{i=0}^{\infty} a_{i}r^{i+2},\ee
which, by comparing the coefficients of like powers of $r$, yields
the recurrence relation
 \be
[(i+\delta)(i+\delta-1)-l(l+1)]a_{i}-2\sum_{n=0}^{i}d_{i-2-n}a_{n}+
2a_{i-2}\lambda=0,\label{reinf}\ee \noindent where $a_{i}=0$ for
$i<0$. Setting $i=0$ in the above relation, we obtain the indicial
equation
 \be [\delta(\delta-1)-l(l+1)-2d_{-2}]a_{0}=0,\label{del}\ee
which is solved by
 \be \delta_{1}={1\over 2}(1-\sqrt{8d_{-2}+(1+2l)^2}),~
\delta_{2}={1\over 2}(1+\sqrt{8d_{-2}+(1+2l)^2}).\label{d}\ee The
Fuchs's theorem \cite {Fuchs} asserts that the generalized series
(\ref{fun}) converges, and both linearly independent solutions of
the Schr\"{o}dinger equation (\ref{ham}) are obtained as generalized
series, one with $\delta=\delta_{1}$ and the other with
$\delta=\delta_{2}$ (except the special case when
$\delta_{2}-\delta_{1}=\sqrt{8d_{-2}+(1+2l)^2}$ is equal to a
non-negative integer). The value of $\delta$ determines the behavior
of $u(r)$ for $r\rightarrow 0$, and only $\delta>{1\over 2}$ is
acceptable \cite{lan}, since only in this case the mean value of the
kinetic energy is finite. Such a solution to the Schr\"{o}dinger
equation (\ref{ham}) exists only if the potential is such that
$d_{-2}> -{1\over 8}$. This solution contains only the series with
$\delta = \delta_{2}$, and in the following will be denoted by \be
u(r,\lambda)=r^{\delta_{2}} \sum_{i=0}^{\infty}
a_{i}r^{i},\label{funtrun}\ee where the dependence on the energy
eigenvalue $\lambda$ is explicitly marked. The coefficients of the
series
 are determined recurrently as \be a_{i}={-2\lambda
a_{i-2}+2\sum_{n=0}^{i-1}d_{i-2-n}a_{n}\over
i(i+\sqrt{8d_{-2}+(1+2l)^2})}.\label{rec}\ee The function
$u(r,\lambda)$ has an important property that $u(0,\lambda)=0$,
which can be regarded as a boundary condition for the radial
equation at $r=0$. The second boundary condition should be chosen
according to the physical bounds in the investigated problem.

\section{Confined potentials}\label{conf}

The simplest  scheme arises in case of radially symmetric potential
$V(r)$ additionally bounded by an infinitely high wall at $r=R$. In
this case the second boundary condition for a particle with angular
momentum $l$ is of the form \be u(R,\lambda)=0,\label{R}\ee yielding
an exact quantization condition that gives the bound state energies
as zeros of a calculable function.  The values of $\lambda$,
determined from the above condition, are will be denoted by
$\lambda_{nl}$, where $n=0,1,...$ counts the number of zeros in the
radial variable in the Sturm-Liouville eigenvalue
problem~\cite{cour}. In the numerical calculation, the function
$u(R,\lambda)$ has to be approximated by truncating the series in
(\ref{funtrun}) at suitably high order $K$, which can be done with
an arbitrary accuracy, as the series is convergent. The truncated
function is a polynomial of degree $K$ in the variable $\lambda$,
that can be used to advantage in determining the numerical values of
its zeros. In this way an increasing amount of bound-state energies
can be obtained with increasing accuracy as the truncation order
$K\rightarrow\infty$. Considering the growing interest in quantum
confined systems, the method (and its possible modifications for
different symmetries of the confining box) can find application in
the studies of semiconductor nanostructures such as quantum
dots~\cite{dot,dot1,dot5}. Here we consider a few examples.

\subsection{Harmonic Oscillator} 
\begin{figure}[h]
\begin{center}
\epsfbox{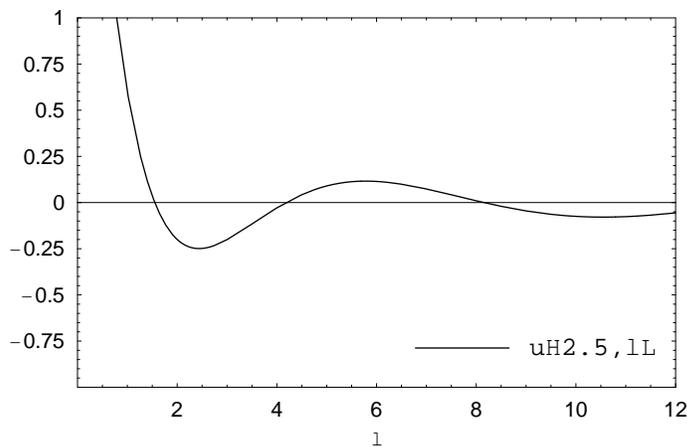}
\end{center}
\caption{\label{fig1:beh} Behavior of $u(R,\lambda)$ for the
spherically symmetric harmonic oscillator with $\omega=1$ and $l=0$
at $R=2.5$. }
\end{figure}

\begin{figure}[h]
\begin{center}
\epsfbox{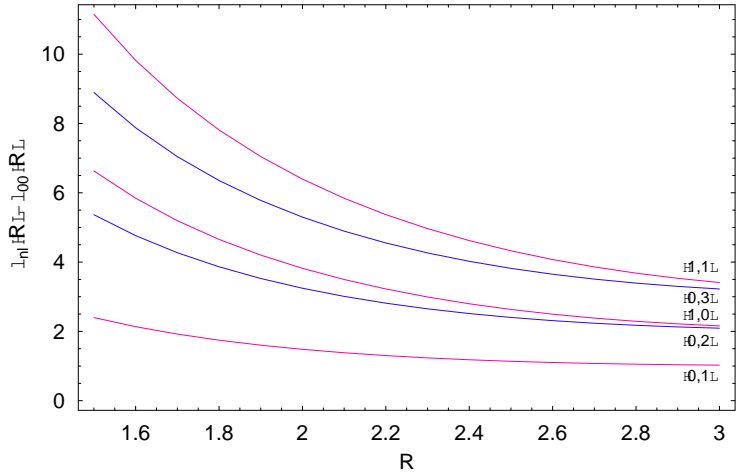}
\end{center}
\caption{\label{fig2:beh} Behavior of
$\lambda_{nl}(R)-\lambda_{00}(R)$ for the harmonic oscillator in
function of the confinement radius $R$.}
\end{figure}

The performance of the FM
will be demonstrated for the spherical harmonic oscillator  \be
V(r)={\omega^2\over 2} r^2,\ee enclosed by a sphere of the radius
$R$. The typical behavior of $u(R,\lambda)$ as a function of
$\lambda$ is shown in Fig.\ref{fig1:beh} in the example of the
oscillator with $\omega=1$ for $l=0$ and $R=2.5$. The approximate
$u(R,\lambda)$ is a polynomial in $\lambda$ and the truncation
affects only its behavior for large $\lambda$. We determine thus the
bound state energies as roots of the obtained polynomial with the
use of NSolve procedure from the Mathematica package, without any
need for introducing starting values for $\lambda$.  The stability
of the numerical results was achieved by increasing the number of
non-vanishing terms $K$ until the values of $\lambda_{nl}(R)$,
corresponding to the states $(n,l)$ of angular momentum $l$, become
stable to the desired accuracy. The values of $\lambda_{nl}(R)$,
obtained for the harmonic oscillator of frequency $\omega=1$, are
compared  in Table \ref{tab:table1} with the eigenvalues of the
unconfined oscillator, $E_{nl}=2n+l+{3\over 2}$. In
Fig.\ref{fig2:beh} the low-lying states energies are plotted as a
function of the confinement radius $R$ . One can observe how the
non-degenerate levels of the confined system approach the
equidistant states of the unconfined oscillator, as the radius of
enclosure grows.

\begin{table}[h]
\caption{\label{tab:table1}The low-lying energy levels of the
harmonic oscillator with $\omega=1$ confined at various~$R$.}
\begin{center}
\begin{tabular}{cccccccccc} \br
$l$& $n$&$\lambda_{nl}(1.5)$ &$\lambda_{nl}(2.5)$ &$\lambda_{nl}(3)$&
$\lambda_{nl}(3.5)$&$\lambda_{nl}(4)$& $E_{nl}$\\
\mr
\hline0&0 &2.5049762&1.5514217&1.5060815&1.5003995&1.5000146&1.5  \\
1&0 &4.9035904&2.6881440&2.5312925&2.5029102&2.5001438&2.5  \\
0&1 &9.1354221&4.1842613&3.6642196&3.5233023&3.5016915&3.5  \\
2&0 &7.8717305&3.9535289&3.5982477&3.5125803&3.5008421&3.5\\
\br
\end{tabular}
\end{center}
\end{table}

\subsection{Anharmonic Oscillator}

The quantum anharmonic oscillator is widely used to describe the
physical phenomena, especially in the condensed matter and molecular
physics. However, the case of a confined anharmonic system has been
discussed only for the one-dimensional example~\cite{bar}. Here we
study the spherically symmetric anharmonic oscillator with a
potential of the form \be V_{AO}(r)= \frac{\omega^2}{2} r^2 +g
r^{2J}\label{anhar}\ee for different values of the power $J$. The
analysis can be simplified by using the rescaled variables
$$\widehat{r}=g^{{1\over 2J+2}} r, \mbox{~~~~and~~~~}  \widehat{\lambda} = \lambda
g^{-{2\over 2J+2}}$$ then Eq.(\ref{ham}) takes a form \be
\left[-{1\over 2 }{d^2\over d\widehat{r}^2}+{l(l+1)\over 2
\widehat{r}^2}+V_{AO}(\widehat{r})\right]u(\widehat{r})=
\widehat{\lambda} u(\widehat{r}),\label{resham}\ee where  \be
V_{AO}( \widehat{r} )=z \widehat{r}^2
+\widehat{r}^{2J}\label{anh}\ee depends on the dimensionless
parameter \be z={\omega^2\over 2}g^{-{4\over 2J+2}},\ee which
accounts for a relative strength of the harmonicity and
anharmonicity. In the following, we skip the hats over $r$ and
$\lambda$. We consider the solution of (\ref{resham}), which is
regular at $r=0$, as given by the generalized power series \be
u(r,\lambda)=r^{l+1}\sum_{i=0}^\infty a_{i}r^{2i}\label{genAO}\ee
with the recurrence relation of the form \be a_{i}={-2\lambda
a_{i-1}+2z a_{i-2}+2a_{i-(J+1)}\over 2i(1+2i+2l)},\ee where
$a_{0}=1$. In the case of the oscillator enclosed by a sphere of the
radius $R$ the bound-state energies are easily determined as zeros
of $u(R,\lambda)$. In Table \ref{tab:table2}
 the numerical results
are presented on the example of the quartic oscillator ($J=2$) with
different values of the parameter $z$ and for different radii of
enclosure $R$. The spectrum of an unconfined anharmonic oscillator
will be studied in the next section.

\begin{table}[h]
\begin{center}
\caption{\label{tab:table2} The bound-state energies
$\lambda_{nl}(R)$ of the confined anharmonic oscillator
}
\lineup
\begin{tabular}{@{}cccccccc}
\br
$l$& $n$& $z$&\0$\lambda_{nl}(0.5)$&$\0\lambda_{nl}(1)$& \0$\lambda_{nl}(1.5)$&
\0$\lambda_{nl}(2)$&\0$\lambda_{nl}(2.5)$ \\
\mr
\hline 0&0 & 5 &20.09875027&\06.39291817&5.10864245&\05.06974045&\05.06966368  \\
 {}&{} & 4 &20.02840681 &\06.13228781&4.69129108&\04.63348473&\04.63330831  \\
 {}&{} & 3 &19.95799389 &\05.86753935&4.24715028&\04.16026790&\04.15984713  \\
 {}&{} & -1 &19.67564622 &\04.76592159&2.12668280&\01.63322595&\01.61278177  \\
{}&{} & -3 &19.53405395 &\04.18860246&0.80447958&-0.38394286&-0.54059117  \\
1&0 & 5 &40.85989845&12.08775101&8.80212012&\08.63744701&\08.63688371\\
{}&{} & 4 &40.76648791&11.73035216&8.16458083&\07.93950715&\07.93833601\\
{}&{} & 3 &40.67302607&11.36972235&7.49829458&\07.18963353&\07.18714592\\
{}&{} & -1 &40.29866554&\09.89468226&4.51245222&\03.40693967&\03.34739553\\
{}&{} & -3 &40.11117731&\09.13754642&2.81077229&\00.78285838&\00.50413322\\
\br
\end{tabular}
\end{center}
\end{table}


\subsection{Hulth\'{e}n potential}
The FM can be also useful for computing energy eigenvalues, in case
when the series expansion of its regular part~(\ref{pot1}) converges
only in a finite interval $\rho_{V}$. In this case, the convergence
of the generalized series solution~(\ref{funtrun}) is granted only 
for $r<\rho_{V}$ and the problem is well defined only for the
confinement radius $R<\rho_{V}$. If the confining box projects
beyond the convergence sphere, we must be very careful, since for
$r>\rho_{V}$ the series representation $u(r,\lambda)$ does not
necessarily coincide with the solution of the radial Schr\"{o}dinger
equation in the potential $V(r)$. As an example, we consider the
Hulth\'{e}n potential \be V(r)={-\delta e^{-\delta r}\over
1-e^{-\delta r}}\label{hult},\ee when the screening parameter
$\delta>0$ is not too large. To employ the FM we expand the
Hulth\'{e}n potential into the Laurent series \be V(r)=-{1\over r}+
V_{reg}(r),\label{hul}\ee with the regular part given by
 \be V_{reg}(r)={\delta\over 2}-\delta
\sum_{n=0}^\infty g_{n} [\delta r]^{2n+1},\label{reghul}\ee where
\be g_{n}={(-1)^n \beta_{n+1}\over [2(n+1)]!},\ee and $\beta_{n}$
are given \cite{Sch} by a convenient expression \be
\beta_{n}=(-1)^n{n\over 2^{2n}-1}\sum _{k=1}^{2n-1}{1\over
2^k}\sum_{j=1}^{k}(-1)^{j} {k \choose j} j^{2n-1}.\ee The
convergence radius of (\ref{reghul}) is $\rho_{V}={2\pi\over
\delta}$, and in the range $0\leq r<\rho_{V}$ the potential
(\ref{hul}) can be approximated with an arbitrary accuracy by a
series with a finite number of terms. With the series truncated
after the $r^{2P+1}$ term, the recursion relation (\ref{rec}) takes
a form \be a_{i}={- 2a_{i-1}+2({\delta \over
2}-\lambda)a_{i-2}-2\sum_{j=1}^{P+1}\delta^{2j}g_{j-1}a_{i-(2j+1)}\over
i(i+2l+1)}.\ee In Table \ref{tab:table3} the numerical results are
presented for different values of $\delta$ at various radii of
enclosure $R<\rho_{V}$. Numerical stability was achieved by
increasing the number of terms both in the series solution
(\ref{funtrun}) and in the potential expansion (\ref{reghul}) until
the approximate values $\lambda_{nl}$ for fixed $R$ become stable to
the quoted accuracy. The table also contains the exact eigenvalues
$E_{nl}$ for the unconfined Hulth\'{e}n potential, obtained
analytically ($l=0$) and by numerical integration ($l\neq
0$)~\cite{integ}.

\begin{table}[h]
\begin{center}
\caption{\label{tab:table3} Bound state energies of the Hulth\'{e}n
potential for various values of the radius of confinement $R$.}
\lineup
\begin{tabular}{@{}cccccccc}
\br
$l$& $n=n_{r}$& $\delta$&\0$\lambda_{nl}(4)$&$\lambda_{nl}(6)$& $\lambda_{nl}(8)$   &$\lambda_{nl}(12)$& $E_{nl}$  \\
\mr
\hline 0&0 & 0.050 &-0.4585448 &-0.4752873&-0.4753117&-0.4753125&-0.4753125  \\
&{}&0.075 & -0.4463941 &-0.4631776&-0.4632024& -0.4632031&-0.4632031 \\
0&1&0.050 & \00.4447886 &-0.0606327&-0.0888523& -0.0975630&-0.1012500\\
{}&{}&0.075 &\00.4567303 &-0.0492481&-0.0776572& -0.0864942&-0.0903125\\
1&0 & 0.050 &\00.1680730&-0.0802458&-0.0947675& -0.0992341&-0.1010425\\
&{}&0.075 &\00.1800057 &-0.0687394& -0.0834014& -0.0879576& -0.0898478  \\
\br
\end{tabular}
\end{center}
\end{table}

\section{Unconfined potentials}\label{meth}

\begin{figure}[h]
\begin{center}
\epsfbox{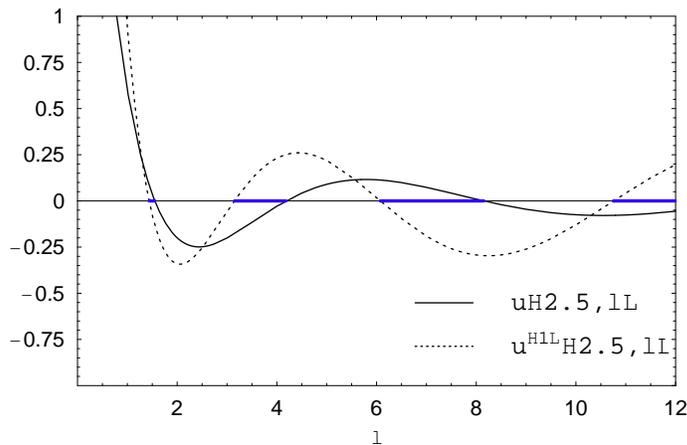}
\end{center}
 \caption{\label{fig3:beh}Behavior of $u(R,\lambda)$
(solid line) and $u^{(1)}(R,\lambda)$ (dashed line) for the
spherically symmetric harmonic oscillator with $\omega=1$ and $l=0$
at $R=2.5$. The three first bounding intervals are marked by thick
line}.
\end{figure}

Now we come to the discussion of unconfined potentials, namely to
the case of a particle with the angular momentum $l$ in a
spherically symmetric potential of the form (\ref{pot}) without any
external enclosure. An unconfined system can be also effectively
treated by the FM, as demonstrated by the calculation of the ground
state energy of the one-dimensional anharmonic oscillator to an
enormous precision of $1184$-digits~\cite{trott}. Here we show that
the method can be implemented in a way that also allows for
computing the excited states energies. We take the generalized
series~(\ref{funtrun}), as a solution of the radial Schr\"{o}dinger
equation, which fulfils the boundary condition $u(\lambda,0)=0$, and
consider two ways of imposing the second boundary condition at
finite $r=R$, namely $u(R,\lambda)=0$ or $u^{(1)}(R,\lambda)=0$. At
$R\rightarrow \infty$ both conditions are satisfied at the same
values of $\lambda$, which correspond to the energy eigenvalues
$E_{nl}$ of the unconfined system, where $n=0,1,2,...$. At finite
$R$ the values of $\lambda$ are different: let us denote by
$\lambda_{kl}$ the values satisfying \be
u(R,\lambda_{kl})=0\mbox{~~for~~} k=0,1,...,\label{u}\ee and by
$\lambda_{kl}^{'}$ those satisfying \be
u^{(1)}(R,\lambda_{kl}^{'})=0, \mbox{~~for~~}
k=0,1,...\label{uprim}\ee In both cases we deal with the
Sturm-Liouville eigenvalue problem; the nodes of the function
$u(r,\lambda_{kl})$ in the radial variable divide thus the domain
$(0,R)$ precisely into $k$-parts~\cite{cour}, and the same is true
for the function $u(r,\lambda_{kl}^{'})$. The energy eigenvalues
satisfy thus the following inequalities:
\be\lambda_{0l}^{'}<\lambda_{0l}<\lambda_{1l}^{'}<\lambda_{1l}<\lambda_{2l}^{'}
<\lambda_{2l}<...<\lambda_{kl}^{'}<\lambda_{kl}<...\label{p}\ee It
is easy to check that the sign of $u(R,\lambda)$ is opposite to that
of $u^{(1)}(R,\lambda)$, for any value $\lambda$ lying within
$(\lambda_{kl}^{'},\lambda_{kl})$, which will be called the bounding
interval in the following. Whereas, for values of $\lambda$ in the
interval $(\lambda_{kl},\lambda_{k+1l}^{'})$ the signs of
$u(R,\lambda)$ and $u^{(1)}(R,\lambda)$ are the same. These
properties are best illustrated on Fig.\ref{fig3:beh}, in the
example of $u(r,\lambda)$ calculated for the spherical harmonic
oscillator with angular momentum $l=0$.

Bound states in the radial potential $V(r)$ are possible only for
$\lambda<\lim_{r\to \infty}V(r)$. In this case, there exists a point
$R_{c}$ such that for $r> R_{c}$ we have $V_{eff}(r,l)-\lambda>0$
and the sign of $u^{(2)}(r,\lambda)$ is the same as that of
$u(r,\lambda)$, which implies that for $r>R_{c}$ the function
$u(r,\lambda)$ must neither have a local maximum if $u(R_{c},\lambda
)>0$, nor a local minimum if $u(R_{c},\lambda)<0$. If both
$u(R_{c},\lambda)$ and $u^{(1)}(R_{c},\lambda)$ are positive
(negative), then $u(r,\lambda)$ and $u^{(1)}(r,\lambda)$ tend
monotonically to the plus (minus) infinity. Therefore, if the point
of imposing the boundary condition,  $R$ is larger than $R_{c}$, the
following inequality is satisfied \be
\lambda_{0l}^{'}<E_{0l}<\lambda_{0l}<...
<\lambda_{kl}^{'}<E_{kl}<\lambda_{kl}<...
<\lambda_{nl}^{'}<E_{nl}<\lambda_{nl},\label{pp}\ee where
$\lambda_{nl} <V_{eff}(R,l)$. For increasing $R$, the energy
$\lambda_{kl}^{'}(R)$ grows and $\lambda_{kl}(R)$ decreases,
approaching the exact eigenvalue $E_{kl}$ from both sides
monotonically. This allows for bounding the energy eigenvalues of an
unconfined system with a required precision. With the approximate
$u(R,\lambda)$, obtained by truncating the number of terms in the
generalized power series to $K$, both (\ref{u}) and (\ref{uprim})
are polynomial equations in $\lambda$. The large set of energy
levels can be thus determined numerically by finding the roots of
polynomials. Generally, the bounding intervals are larger for higher
states but for increasing $R$ all the bounding interval shrink. We
determine thus the new bounding intervals with the value of $R$
increased by a suitably chosen $\Delta R$. The iteration procedure
is repeated for $R_{i}=R+i \Delta R$, until the bounding energies
for a chosen state $(k,l)$ become equal to the accuracy desired,
which determines its energy with that accuracy. If the procedure
does not converge for the state of interest, the number of terms
$K$, which are included in the power series (\ref{funtrun}) should
be increased.

\subsection{The Kratzer  potential}

\begin{figure}[h]
\begin{center}
\epsfbox{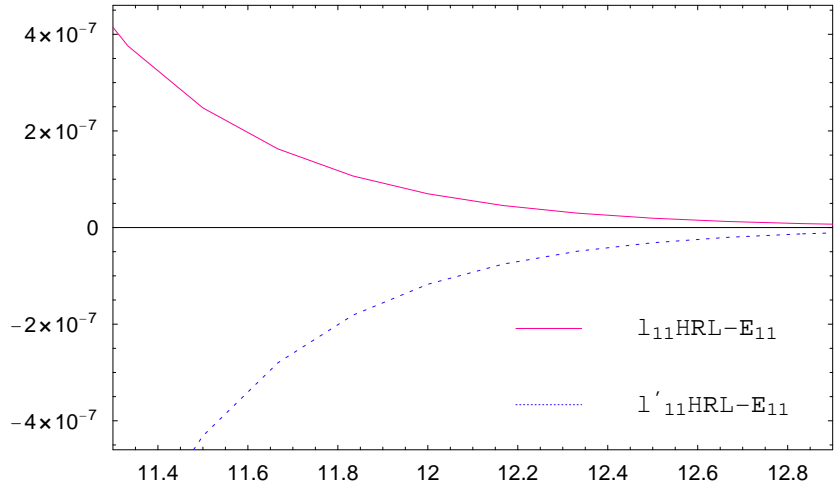}
\end{center} \caption{\label{fig4:beh}Convergence
of $\lambda_{11}(R)-E_{11}$, and $\lambda_{11}^{'}(R)-E_{11}$ to
zero, demonstrated in function of $R$}
\end{figure}

For demonstrating the convergence of the algorithm previously
formulated, we first consider the Schr\"{o}dinger problem in the
Kratzer potential \be V(r)={d_{-2}\over r^2}+{d_{-1}\over
r},\label{krat}\ee for which the exact energy levels are given by
\be E_{nl}=-2 d_{-1}^2(2
n+1+\sqrt{(2l+1)^2+8d_{-2}})^{-2},(n=0,1,2,...)\ee Choosing
$d_{-1}=-8$, and $d_{-2}=4$ as the parameters of the potential
(\ref{krat}), we carry out the calculation for the state $(1,1)$,
taking the number of terms in the power series (\ref{funtrun})
suitably large ($K=160$) in order to assure the numerical stability.
The Table \ref{tab:table4} shows with $14-$digit precision the
values of $\lambda_{11}(R_{i})$, obtained from the condition
(\ref{u}), and those of $\lambda_{11}^{'}(R_{i})$, obtained from
(\ref{uprim}), for $R_{i}=R+i \Delta R$. The bounding interval
shrinks very fast and the exact value of energy is easily determined
with 14-digit precision, $E_{11}=-1.4476568219254$. Figure
\ref{fig4:beh} shows how the deviations from the exact energy
monotonically diminish for increasing $R$, the energy difference
$\lambda_{11}(R)-E_{11}$ approaches zero from above, and
$\lambda_{11}^{'}(R)-E_{11}$ from below.
\begin{table}[h]
 \caption{\label{tab:table4} Comparison of the convergence of
$\lambda_{11}(R)$
 and $\lambda_{11}^{'}(R)$ to the
exact value $E_{11}=-1.4476568219254$ for different values of $R$
and $\Delta R$}

\begin{center}
\begin{tabular}{ccccc}
\br
   $i$ & $R$& $\Delta R$&$\lambda_{11}(R_{i})$ &$\lambda_{11}^{'}(R_{i})$  \\
\mr

\hline 0 &{5} &{0.5}& -1.3178526137388&-1.8092879070838  \\
1 &{} &{}& -1.3839260262988&-1.6834926644347  \\
2 &{} &{}& -\underline{1.4}178923369387&-1.5810678698537  \\
3&{} &{}& -\underline{1.4}345142511694&-1.5094745737014  \\
\hline 0 &{8} &{1}&  -\underline{1.44}68612319721&-\underline{1.4}500438665400  \\
 1  &{} &{}&-\underline{1.447}5622942994&-\underline{1.447}8789515475  \\
 2  &{} &{}&-\underline{1.4476}472391241&-\underline{1.4476}761151204 \\
 3&{} &{}&-\underline{1.44765}59628487&-\underline{1.44765}83796562 \\
\hline 0 &14 & {2}&  -\underline{1.44765682}15565&-\underline{1.44765682}24842 \\
1 &{} & {}&  -\underline{1.44765682192}38&-\underline{1.44765682192}76 \\
2 &{} & {}&  -\underline{1.4476568219254}&-\underline{1.4476568219254} \\
\mr
\end{tabular}
\end{center}
\end{table}

\newpage

\subsection{Anharmonic oscillator}

The unconfined quantum anharmonic oscillator is one of the most
frequently discussed quantum systems. Especially the one-dimensional
example, being the simplest tractable model of quantum field theory
\cite{BW}, is routinely used for examining the validity of various
approximation methods \cite{ao2}, as its exact solution can be
numerically determined to an arbitrary accuracy
\cite{trott,PhRep,ao1,Gomez}. The $D-$dimensional case is much less
studied, generally, a spherically symmetric anharmonic potential
(\ref{anhar}) for different values of the anharmonicity power $J$ is
discussed. The quartic potential ($J=2$) was studied by means of
various approximation methods, e.g. the self-similar approximation
\cite{aosph1},
the random phase approximation \cite{rpa}, and the artificial
perturbation method~\cite{aosph3}. There are also a few reports in
the literature of the numerically exact results for the quartic
\cite{PhRep,witwit,aosph2}, sextic ($J=3$) \cite{chaud,witwit}, and
octic ($J=4$) potential \cite{witwit} in the limited range of the
anharmonicity parameter $z$.

Here we show that the FM enables us to determine effectively the
spectrum of the unconfined spherical anharmonic
oscillator~(\ref{anhar}) in the wide range of the parameter $z$.
With the solution of the Schr\"{o}dinger equation $u(r,\lambda)$ in
the form of the generalized series~(\ref{genAO}) we determine the
numerical values of bound-state energies, using the procedure
formulated in the beginning of this section. After checking that the
results available in the literature \cite{chaud,witwit,aosph2} are
easily recovered to the quoted accuracy, we performed an extensive
calculation of the spectrum of spherically symmetric anharmonic
oscillators. The highly accurate results presented here may serve
for testing the quality of various approximation methods. Especially
challenging test is provided by the data obtained for negative
values of the parameter $z$, when the anharmonic potential has a
Mexican hat shape. This range was not explored before, and the
numerical data for bound-state energies were lacking. Therefore, in
table \ref{tab:table5} we present our results for several lowest
states $(n,l)$ energies of the quartic oscillator ($J=2$) at
negative values of $z$. In table \ref{tab:table6} we compare the
bound-state energies for oscillators with different anharmonicity
powers ($J=2,3,4$) at various values of $z$. The results up to
8-decimal precision are quoted, but it is not difficult to improve
the accuracy at will. However, one has to note that the effort of
the calculation increases for higher states. The appropriate
shrinking of the intervals bounding the higher states energies is
achieved only at large $R$, which requires the larger number of
terms $K$ to be included in the power series. The computational
effort increases strongly, when the parameter $z$ becomes more
negative, i.e. for increasing radius of the hat. For example,
equality of the values $\lambda_{00}(R)$ and $\lambda_{00}^{'}(R)$
with 8-digit accuracy was achieved for $z=-2$ at the radius $R=3$,
which requires $K=65$, while for $z=-10$ the same is obtained only
at $R=3.9$, which requires $K=140$ .

\begin{table}[h]
\caption{\label{tab:table5} The lowest bound state energies
 of the quartic oscillator for negative values of $z$. }
\begin{center}
\lineup
\begin{tabular}{cccccccc}
\br
$z$ &$\0E _{00}$& $\0E _{01}$& $\0E _{02}$ &$\0E _{03}$ &$\0E _{04}$ \\
\mr \hline
-10& -21.88965823&-21.66759959&-21.22775192&-20.57776933&-19.72742347\\
-9& -17.30811508&-17.05634313&-16.56006280&-15.83155272&-14.88543607\\
-8& -13.23757818&-12.94630409&-12.37723745&-11.55086571&-10.48917435\\
-7& \0-9.68064055&\0-9.33403278&\0-8.66866023&\0-7.71939851&\0-6.51858369\\
-6& \0-6.64062824&\0-6.21133359&\0-5.41635974&\0-4.31340468&\0-2.94717508\\
\mr
\end{tabular}
\end{center}
\end{table}

\newpage

\begin{table}[h]
\caption{\label{tab:table6} The low-lying eigenvalues of anharmonic
oscillators for certain powers of anharmonicity $2J$ at different
values of the parameter $z$} \lineup
\begin{center}
\begin{tabular}{cccccccc}
\br
$J$ &$z$&\0$E_{00}$&\0$E_{01}$& \0$E_{02}$ &\0$E_{10}$ &\0$E_{03}$ \\
\mr \hline
2&-5 &-4.11911133&-3.55958100&-2.59137576&-0.21067602&-1.30027463\\
&-4 &-2.10349462&-1.34175838&-0.15162331&\01.57335020&\01.36098787\\
&-3 &-0.54212526&\00.50094741&\01.95580851&\03.17712298&\03.71671698\\
&-2 &\00.66142890&\02.04064501&\03.78779723&\04.66539082&\05.81395433\\
&0.1&\02.46463653&\04.58321826&\06.96529241&\07.45891176&\09.56314673\\
&0.5&\02.73789227&\04.99143053&\07.49177505&\07.94240398&10.19704822\\
&1&  \03.05794573&\05.47591999&\08.12171427&\08.52673739&10.95972825\\
&2& \03.63948205&\06.37163094&\09.29940583&\09.63362791&12.39687279\\
&5&\05.06966367 &\08.63688366 &12.33685580&12.55011513&16.15843962\\
&8&\06.21722563 &10.49477725
&14.87064591&15.02454165&19.33921639\\\hline
3&-8& -5.00982691&-4.08581639&-2.49927339&\01.27019537&-0.37612149\\
&-5& -1.39861613&-0.06070241&\01.93123653&\04.23192390&\04.43629185\\
&-2& \01.24144988&\03.20845649&\05.72356878&\07.05803262&\08.68689446\\
& -1&\01.94504486&\04.15238570&\06.86055253&\07.97745578&\09.98954703\\
&0.1&\02.63985266&\05.11853012&\08.04543812&\08.96977843&11.36185573\\
&1
&\03.15630057&\05.85836853&\08.96728095&\09.76393591&12.44002444\\\hline
4&-8& -3.20712029&-1.83434272&\00.36915524&\04.03170882&\03.24522274\\
& -5&-0.57239228&\01.19930711&\03.74932588&\06.24632114&\06.93207343\\
& -2&\01.55575737&\03.85587069&\06.82325601&\08.47500053&10.35857448\\
&-1&\02.16525909&\04.66276445&\07.78250948&\09.21918447&11.44445127\\
&0.1& \02.78594828&\05.50821786&\08.80159788&10.03495445&12.60737664\\
&
1&\03.25887003&\06.16887293&\09.60819432&10.69825337&13.53482309\\\hline
5& 0.1&\02.91188167&\05.81747881&\09.38031919&10.83477815&13.54422429\\
& 1&\03.35601445&\06.42909865&10.11830492&11.42342067&14.38432780\\
 \mr
\end{tabular}
\end{center}
\end{table}
\begin{figure}[h]
\begin{center}
\epsfbox{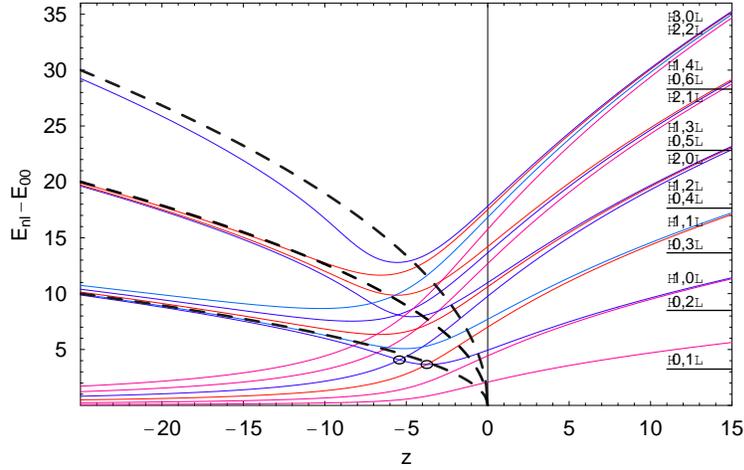}
\end{center} \caption{\label{fig5:beh}The behavior $E _{nl}-E _{00}$
in function of $z$, the asymptotic behavior of
$E_{n}-E_{0}=n\sqrt{-4z}$ is shown by dashed lines. Two points of
level crossing are marked by circles.}
\end{figure}

Our results for the quartic oscillator are presented graphically in
Fig.\ref{fig5:beh}. The excitation energies with respect to the
ground state, $E_{nl}-E_{00}$, are plotted in function of the
parameter $z$, which covers the range $-15<z<25$. The values of
$z<0$ correspond to the Mexican hat, and those of $z>0$ to the
single-well shape of the anharmonic potential (\ref{anh}). It is
interesting to note that, in spite of this difference, the
dependence of excitation energies on $z$ is smooth at the point
separating the two cases, $z=0$, which corresponds to the strong
coupling limit, $g\rightarrow \infty$. Instead, the behavior of
excitation energies in the weak coupling limit ($|z| \rightarrow
\infty$) is very different for positive and negative $z$. For $z
{\rightarrow}{\infty}$ we recover the radial harmonic oscillator of
the frequency $2z$, which energy eigenvalues are given by \be E_{nl}
\begin{array}{c}
   \\
  \longrightarrow \\
  z\rightarrow\infty \\
\end{array}
(2n+l+{3\over 2})\sqrt{2z}.\ee In Fig.\ref{fig5:beh} one can observe
how the levels with the same $2n+l$ become degenerate for increasing
$z$. For $z<0$ the Mexican hat shaped anharmonic potential has a
minimum at $r_{min}=\sqrt{-z\over 2}$. If $z^3>>l^2$, the minimum of
$V_{eff}$ is close to $r_{min}$ and the effective potential is
approximated well by  \be V_{eff}(r)\approx -{z^2\over 4}+{1\over
2}(-4z)(r-\sqrt{{-z\over 2}})^2\label{apph},\ee which does not
depend on $l$, and corresponds to the one-dimensional harmonic
oscillator of the frequency $\sqrt{-4z}$.
Therefore, in the limit $z {\rightarrow}{-\infty}$ we have
 \be E_{n}
\begin{array}{c}
   \\
  \longrightarrow \\
  z\rightarrow - \infty \\
\end{array}
-{z^2\over 4}+(n+{1\over 2})\sqrt{-4z}(n=0,1,2,...)\ee In
Fig.\ref{fig5:beh} the grouping of states with different quantum
numbers $l$ but the same value of $n$ can be observed for strongly
negative values of $z$. For $z\rightarrow - \infty$ all the states
in the group approach the asymptotic behavior of
$E_{n}-E_{0}=n\sqrt{-4z}$.

One can also notice an interesting phenomenon of level crossing that
appears in the range of $z<0$: at certain negative value of $z$ two
adjacent eigenvalues become degenerate. The two examples in
Fig.\ref{fig5:beh} are marked by circles: the crossing point for the
states $(1,0),(0,3)$, which appears at $z_{1}\cong-3.73656382$, and
the crossing point for $(1,0),(0,4)$, which appears at
$z_{2}\cong-5.42007803$. The configuration of five lowest states,
which for $z>z_{1}$ is given by~$ (0,0),(0,1),(0,2),(1,0),(0,3)$,
changes into ~$(0,0),(0,1),(0,2),(0,3),(1,0)$ for $z_{2}<z<z_{1}$,
and into $(0,0),(0,1),(0,2),(0,3),(0,4)$ for $z<z_{2}$. It should be
stressed that in the case of one-dimensional anharmonic oscillator
the phenomenon of level crossing does not appear, this becomes
possible only for systems of $D\geq 2$ dimensions.

\section{Conclusion }

We have shown that the application of the Fr\"{o}benius method to
the spherically symmetrical potentials of the form
$V(r)={d_{-2}\over r^2}+{d_{-1}\over r}+\sum_{i=0}^{\infty}
d_{i}r^i$ allows an easy determination of the energy spectrum. This
was demonstrated first for systems enclosed in a spherical box of
the radius $R$, by studying the confined harmonic and anharmonic
oscillators and Hulth\'{e}n potential. With the increasing radius of
confinement the bound states energies have been shown to approach
those of the corresponding unconfined systems. Even in the case of
 the Hulth\'{e}n potential, when the convergence radius of the potential
 is finite, we obtain quite good approximations to the low-lying spectrum
of the unconfined potential, if the screening is not too strong.
Later, we developed an efficient scheme for computing eigenvalues of
the Schr\"{o}dinger equation for unconfined potentials with a
controlled accuracy. The method allowed us to determine the
low-lying states energies of spherically symmetric anharmonic
oscillators with very high accuracy and moderate effort.
Determination of energies becomes computationally more demanding for
higher states, since more terms have to be included in the
generalized power series. Our calculations cover a broad range of
anharmonic parameters, both in the case of single well potential and
for the Mexican hat shape. In the later case the computational
effort increases strongly with the increasing radius of the hat.

The method presented in this work can be easily applied for
computing the precise spectrum of other spherically symmetric
potentials. In the present work, we have studied the case of
three-dimensional space but the calculation of energy levels in the
two-dimensional case can be performed along similar lines. The
results in even and odd higher dimension $D$ can be easily derived
from those in $2-$ and $3-$dimensional space, respectively, via the
transformation $l\rightarrow l+\frac{D-3}{2}$. One has to add that
the method can be also used to determine the approximate wave
functions. After substituting the calculated bound-state energy to
the recursion relation~(\ref{rec}) the coefficients of the
generalized series~(\ref{funtrun}) can be successively determined in
order to obtain the unnormalized wave function as a sum of the
series.

\end{document}